\documentclass[preprint,review,12pt,sort&compress]{./template/elsarticle/elsarticle}
\usepackage{epstopdf}
\usepackage{amsmath}
\usepackage{xcolor}
\usepackage{lineno,hyperref}
\usepackage{placeins} 
\usepackage{array} 
\usepackage{mathtools}

\modulolinenumbers[5]

\pdfoutput=1

\journal{Combustion and Flame}
\newcommand*\diff{\mathop{}\!\mathrm{d}} 
\newcommand{\sm}[1]{{\scriptscriptstyle#1}} 
\newcolumntype{P}[1]{>{\centering\arraybackslash}p{#1}}

\begin{document}

\begin{frontmatter}

\title{\textcolor{black}{DNS Study of the Global Heat Release Rate During Early Flame Kernel Development under Engine Conditions}}


\author[mymainaddress]{Tobias Falkenstein}

\author[mysecondaryaddress]{Seongwon Kang}

\author[mymainaddress]{Liming Cai}

\author[mymainaddress]{Mathis Bode}

\author[mymainaddress]{Heinz Pitsch\corref{mycorrespondingauthor}}
\cortext[mycorrespondingauthor]{Corresponding author}
\ead{office@itv.rwth-aachen.de}

\address[mymainaddress]{Institute for Combustion Technology, RWTH Aachen University, 52056 Aachen, Germany}
\address[mysecondaryaddress]{Department of Mechanical Engineering, Sogang University, Seoul 121-742, Republic of Korea}

\begin{abstract}
\textcolor{black}{Despite the high technical relevance of early flame kernel development for the reduction of cycle-to-cycle variations in spark ignition engines, there is still a need for a better fundamental understanding of the governing in-cylinder phenomena in order to enable resilient early flame growth. To isolate the effects of small- and large-scale turbulent flow motion on the young flame kernel, a three-dimensional DNS database has been designed to be representative for engine part load conditions. The analysis is focussed on flame displacement speed and flame area in order to investigate effects of flame structure and flame geometry on the global burning rate evolution. It is shown that despite a Karlovitz number of up to~13, which is at the upper range of conventional engine operation, thickening of the averaged flame structure by small-scale turbulent mixing is not observed. After ignition effects have decayed, the flame normal displacement speed recovers the behavior of a laminar unstretched premixed flame under the considered unity-Lewis-number conditions. Run-to-run variations in the global heat release rate are shown to be primarily caused by flame kernel area dynamics.
The analysis of the flame area balance equation shows that turbulence causes stochastic flame kernel area growth by affecting the curvature evolution, rather than by inducing variations in total flame area production by strain. Further, it is shown that in local segments of a fully-developed planar flame with similar surface area as the investigated flame kernels, temporal variations in flame area rate-of-change occur. Contrasting to early flame kernels, these effects can be exclusively attributed to curvature variations in negatively curved flame regions.}
%
\end{abstract}


\end{frontmatter}

\section{Introduction}
\label{sec:intro}
%
\textcolor{black}{Combustion stability  is a prerequisite for more efficient spark ignition (SI) engine operation~\cite{Aleiferis04_lean_ccv,Jung17_lean_si_ccv} and reduced engine-out emissions~\cite{Milkins74_CO_HC_ccv_exp,Karvountzis17_emissions_ccv}. However, the reduction of cycle-to-cycle variations~(CCV) by design or control measures is challenging due to the complex multi-physics and multi-scale nature of the in-cylinder processes, which are not completely understood. }
From experimental investigations, it is known that~CCV are correlated with fluctuations in early flame kernel development~\cite{Schiffmann17_exp}. Even under conditions that lead to misfires, spark ignition was found to successfully initiate a flame kernel, which extinguishes during the early growth phase~\cite{Peterson11_exp_misfire}. By using advanced optical diagnostics, prolonged chemical time scales due to exposure of the small flame to locally lean or diluted mixture~\cite{Peterson14_early_flame_exp}, or fluctuations in large-scale flow structures~\cite{Bode17_piv_early_kernel,Zeng18_exp} and kinetic energy~\cite{Buschbeck12_piv_kernel} could be identified as sources for cyclic variability. Under fully premixed in-cylinder conditions, variations in burning velocity during the transition phase from ignition kernels to fully developed turbulent flames were found to be one major contributor to CCV~\cite{Zeng18_exp}. However, the role of turbulence during early flame kernel development 
has not been assessed experimentally \textcolor{black}{due to diagnostic challenges. To isolate certain physical effects that govern early flame growth, well-designed numerical experiments in simplified configurations can be very useful.} \par
\textcolor{black}{Various numerical studies on flame kernel development in simplified configurations are available in the literature, which are mainly based on} either two-dimensional DNS with complex transport and chemistry or three-dimensional DNS with simplified transport and/or chemistry. 
Fundamental investigations on laminar flame kernel/flow interactions were performed by Echekki and Kolera-Gokula~\cite{Echekki07_kernel_vortex_map_01} and Vasudeo et al.~\cite{Vasudeo10_kernel_vortex_map_01}. Two governing physical parameters, the ratio of vortex and flame sizes, as well as the ratio of vortex strength and burning velocity were identified, and a qualitative regime diagram was proposed. A similar study on flame kernel development, but in two-dimensional turbulence, was conducted by Reddy and Abraham~\cite{Reddy11_regime_map}. In the proposed regime diagram, the ratio of integral length scale and kernel size, as well as the ratio of turbulent velocity fluctuation and burning velocity appear as independent parameters. While these studies have established a qualitative understanding of global flame kernel/flow interactions, quantitative differences between identified regimes and consequences for cyclic variations in engines could not be extracted from the datasets. \par 
Flame propagation characteristics of small flame kernels developing in more realistic turbulence have been analyzed in several DNS studies. Klein et al.~\cite{Klein06_kernel_dns_stretch} investigated response of the flame displacement speed to curvature and found significant differences as compared to a statistically planar flame. The flame structure of an early kernel in stratified mixture was studied by Chakraborty et al.~\cite{Chakraborty07_spark_ign_turb_dns}. Increased turbulent velocity fluctuations were found to have detrimental effects on ignition due to increased mixture fraction gradients and locally reduced flame propagation velocities. From a parametric study on flame kernel development at various equivalence ratios and turbulence intensities~\cite{Fru11_kernel_dns_3d}, flame/flame interactions and local extinction were found to become increasingly important at higher velocity fluctuations. Extinction of an ignition kernel before reaching the critical radius for self-sustaining flame propagation, was studied by Uranakara et al.~\cite{Uranakara17_ign_kernel_3d}. A particle-based analysis was utilized to understand the effect of turbulence-enhanced heat loss leading to global extinction. \par

Despite the numerous existing studies on various aspects of early flame kernel development, it is still uncertain if the known phenomena can sufficiently explain the occurrence of~CCV in~SI engines, and which effects are dominant. In the present work, a DNS database has been carefully designed with all parameters being representative for engine part load conditions to allow for conclusions of practical relevance. A systematic parameter study was carried out to isolate the consequences of a particular feature of early kernel growth in engines, viz. the initially small flame size relative to the integral length scale of turbulence. \textcolor{black}{Here, flame kernel/turbulence interactions are considered in the unity-Lewis-number limit to establish a well-defined reference case, which is a prerequisite to isolate the substantial effects of differential diffusion under otherwise identical conditions~\cite{Falkenstein19_kernel_Le_I_cnf,Falkenstein19_kernel_Le_II_cnf}.} \par
The present manuscript is organized as follows. In Sect.~\ref{sec:math_form}, the governing parameters for the global heat release rate are derived. Details of the DNS datasets are described in Sect.~\ref{sec:dns_data_base} \textcolor{black}{and the relevance for actual engines is discussed}. A comprehensive analysis of the integral heat release rate during flame development \textcolor{black}{with consideration of both flame structure and flame geometry evolution} is presented in Sect.~\ref{sec:react_flow}.
\section{Mathematical Formulation: Key Quantities of Interest}
\label{sec:math_form}
Technical devices for chemical energy conversion are typically characterized by quantities which describe the overall system behavior. \textcolor{black}{One of the most important characteristics impacting SI engine thermal efficiency is the global heat release rate}, which would ideally approach the limit of constant-volume combustion. Hence, integral reaction progress is one of the main quantities to be considered for engine combustion process assessment and combustion model accuracy tests. In the following, the key flame parameters which will be subject of detailed analysis in Sect.~\ref{sec:react_flow}, will be derived.\par
To quantify the global heat release rate, a reaction progress variable~$c$ is considered, which is commonly defined as the sum of mass fractions of major combustion products~\cite{Pitsch06_ann_review}. Applying the Reynolds transport theorem to the transport equation of~$c$ yields its rate of change inside the domain~$\Omega$ encompassing the entire flame:
\begin{eqnarray}
\frac{D \left. \overline{ \rho c }\right|_{\Omega}}{D t} & = &
\frac{1}{V_{\Omega}} \int_{\Omega}\left[ \frac{\partial \left(\rho c \right)}{\partial t} + \frac{\partial \left( \rho u_i c \right)}{\partial x_i}\right] \diff V \nonumber\\ 
& = &
\frac{1}{V_{\Omega}} \int_{\Omega}\left[ \frac{\partial }{\partial x_j}\left(\rho D_{\mathrm{th}} \frac{\partial c}{\partial x_j} \right) + \dot{\omega}_c\right] \diff V  \nonumber\\
& = &
\left.\overline{\dot{\omega}_c}\right|_{\Omega}.
\label{eq:math_form_global_srcPROG}
\end{eqnarray}
Here, $V$ denotes a volume, $\rho$ is the density, $u_i$ is the  local fluid velocity, $\dot{\omega}_c$ is a chemical source term and $D_{\mathrm{th}}$ is the thermal diffusion coefficient. For simplicity, unity Lewis numbers are assumed for all species. Note that the diffusive fluxes at the domain boundaries are considered to be zero, i.e.\ the volume integral of the diffusive term over~$\Omega$ in Eq.~(\ref{eq:math_form_global_srcPROG}) vanishes according to Gauss's Theorem. \par
\textcolor{black}{Eq.~(\ref{eq:math_form_global_srcPROG}) may be converted into a kinematic formulation~\cite{Boger98_gen_fsd} valid at every iso-level of~$c$, by introducing the flame displacement speed~$\mathrm{s}_{\mathrm{d}}$:}
\begin{align}
\mathrm{s}_{\mathrm{d}} & = \frac{1}{\rho \left|\nabla c  \right|} \left( \dot{\;\omega_{c}} - \frac{\partial}{\partial x_n} \left( \rho D_{\mathrm{th}} \left|\nabla c \right| \right) \right)
- D_{\mathrm{th}}\kappa \nonumber \\
& = \mathrm{s}_{\mathrm{rn}} - D_{\mathrm{th}}\kappa .
\label{eq:sd_thin_rct_zones}
\end{align} 
\textcolor{black}{Note that the diffusion term has been split into contributions by normal and tangential diffusion~\cite{Echekki99_diff_term_split}, which can be useful to express~$\mathrm{s}_{\mathrm{d}}$ by a normal-propagation velocity~$\mathrm{s}_{\mathrm{rn}}$, and a curvature term. The resulting propagation form of the scalar transport equation reads}
\begin{align}
\frac{D \left. \overline{ \rho c }\right|_{\Omega}}{D t} & = 
\frac{1}{V_{\Omega}} \int_{\Omega}\left( \rho {\mathrm{s}}_{\mathrm{d}} \left|\nabla c \right| \right) \diff V \nonumber\\
& = 
\left.\overline{\rho {\mathrm{s}}_{\mathrm{d}} \left|\nabla c \right|}\right|_{\Omega} = \frac{\left.\overline{\rho {\mathrm{s}}_{\mathrm{d}} \left|\nabla c \right|}\right|_{\Omega}}{\left.\overline{ \left|\nabla c \right|}\right|_{\Omega}} \left.\overline{ \left|\nabla c \right|}\right|_{\Omega}  \nonumber\\
& = \left< \rho {\mathrm{s}}_{\mathrm{d}} \right>_{\mathrm{s,}\Omega} \overline{\Sigma}_{\Omega} \nonumber  \\
& = 
 \left(\rho_{\mathrm{u}}\mathrm{s}_{\mathrm{l,u}}^{\sm{0}}\right) \overline{\mathrm{I}}_0 \; \overline{\Sigma}_{\Omega},
%
\label{eq:math_form_global_srcPROG_prop}
\end{align}
where the generalized flame surface density~$\overline{\Sigma}$ is computed by averaging the flame surface densities corresponding to every iso-level $c^*$~\cite{Boger98_gen_fsd}:
\begin{align}
\overline{\Sigma}_{\Omega} & =  \int_0^1 \Sigma_{\Omega}^* \diff c^* \nonumber \\
& =  \int_0^1 \int_{\Omega} F \delta \left(c-c^*\right) \left|\nabla c \right| \diff V \diff c^*
=  \left.\overline{ \left|\nabla c \right|}\right|_{\Omega} \nonumber \\
& = \frac{1}{V_{\Omega}} \int_0^1  A_{c=c^*} \diff c^* = \frac{\overline{A}_{c,\Omega}}{V_{\Omega}},
\label{eq:gen_fsd}
\end{align}
where $A_c$ denotes the area of a progress variable iso-surface and it is assumed that the filter operator $F$ is given by a box filter in physical space. Note that $\overline{\Sigma}_{\Omega}$ can be interpreted as a representative surface area per unit volume, obtained by averaging all scalar iso-surface areas in that volume.\par
Similarly, the generalized iso-surface average of any quantity $\vartheta$ is introduced:
\begin{align}
\left< \vartheta \right>_{\mathrm{s,}\Omega}
& =  \frac{\int_0^1 \int_{\Omega} F \vartheta \delta \left(c-c^*\right) \left|\nabla c \right| \diff V \diff c^*}
{\int_0^1 \int_{\Omega} F \delta \left(c-c^*\right) \left|\nabla c \right| \diff V \diff c^*} \nonumber \\
& =  \frac{\left.\overline{\vartheta \left|\nabla c \right|}\right|_{\Omega}}{\left.\overline{ \left|\nabla c \right|}\right|_{\Omega}},
\label{eq:gen_fsd_avg}
\end{align}
i.e.\ the averaging operation is defined by computing the surface integral of $\vartheta$ over each scalar iso-surface, averaging over all iso-levels, and normalizing by the averaged scalar iso-surface area. \par
In Eq.~(\ref{eq:math_form_global_srcPROG_prop}), the global stretch factor is defined by:
\begin{equation}
\overline{\mathrm{I}}_0 = \frac{\left< \rho {\mathrm{s}}_{\mathrm{d}} \right>_{\mathrm{s,}\Omega}}{\rho_{\mathrm{u}}\mathrm{s}_{\mathrm{l,u}}^{\sm{0}}},
\label{eq:I0_def_01}
\end{equation}
where $\rho_{\mathrm{u}}$ and $\mathrm{s}_{\mathrm{l,u}}^{\sm{0}}$ are the density and laminar burning velocity with respect to the unburned mixture of an unstretched, premixed flame. Originally, $\mathrm{I}_0$ was introduced for modeling of stretched flames~\cite{Bray90_stretch_factor}. In general, $\mathrm{I}_0$ quantifies the deviation from the reaction-diffusion balance of a reference flame, e.g.\ due to spark ignition, stretch, or turbulent mixing effects. \par
Eqs.~(\ref{eq:math_form_global_srcPROG}) and~(\ref{eq:math_form_global_srcPROG_prop}) can be combined to express the integral reaction progress variable rate-of-change as function of the global stretch factor, and the average flame area per domain volume:
\begin{equation}
\frac{D \left. \overline{ \rho c }\right|_{\Omega}}{D t}  = \left.\overline{\dot{\omega}_c}\right|_{\Omega} =  
\frac{\rho_{\mathrm{u}}\mathrm{s}_{\mathrm{l,u}}^{\sm{0}}}{V_{\Omega}} \cdot \overline{\mathrm{I}}_0 \cdot
 \overline{A}_{c,\Omega}.
\label{eq:key_params}
\end{equation}
\textcolor{black}{Note that~$\overline{\mathrm{I}}_0$ and~$\overline{A}_{c,\Omega}$ are measures for externally invoked changes in flame structure and geometry, respectively}. In this work, analysis of flame kernel development aims at identifying cause-effect-relationships that determine the temporal evolution of these two key parameters $\overline{\mathrm{I}}_0$ (cf.\ Sect.~\ref{ssec:mean_Sd}) and $\overline{A}_{c,\Omega}$ (cf.\ Sect.~\ref{ssec:flame_area}), as well as their dynamics. \par
\section{DNS Database}
\label{sec:dns_data_base}
To approach the complexity of early flame kernel development and allow for conclusive analyses, DNS of three flame configurations under identical conditions, but with different initial flame geometry have been computed. In the following, the equations and solution procedures utilized in the DNS are briefly presented (cf.\ Sect.~\ref{ssec_govern_eq} and~\ref{ssec:ciao_discr}). Details on the flow and flame conditions, as well as on flame initialization, are provided in Sects.~\ref{ssec:phys_config} and~\ref{ssec:ign}. \textcolor{black}{Finally, the approximations and assumptions introduced to realize an engine-relevant DNS setup are discussed in Sect.~\ref{sec:engine_relevance}.}

\subsection{DNS System of Equations}
\label{ssec_govern_eq}
The dataset is based on the solution of reacting flow equations for constant-volume combustion in the low-Mach number limit. Mass conservation is expressed by
\begin{equation}
\frac{\partial{\rho}}{\partial{t}} + \frac{\partial{\left(\rho u_j\right)}}{\partial{x_j}} = 0.
\label{eq:continuity}
\end{equation}
The balance equation for fluid momentum reads
\begin{equation}
\frac{\partial{\left(\rho u_i\right)}}{\partial{t}} + \frac{\partial{\left( \rho u_j u_i \right)}}{\partial{x_j}} = -\frac{\partial{p^{\left(2\right)}}}{\partial{x_i}} + \frac{\partial{\tau_{ij}}}{\partial{x_j}},
\label{eq:momentum}
\end{equation}
where the viscous stress tensor is given by
\begin{equation}
\tau_{ij} =  2 \mu \left( S_{ij}  - \frac{1}{3} \delta_{ij} \frac{\partial{u_k}}{\partial{x_k}}  \right),
\label{eq:visc_stress_tensor}
\end{equation}
and $\mu$ is the dynamic viscosity. The strain rate tensor $S_{ij}$ is defined as:
\begin{equation}
S_{ij} = \frac{1}{2} \left( \frac{\partial{u_i}}{\partial{x_j}}  + \frac{\partial{u_j}}{\partial{x_i}} \right).
\label{eq:strain_rate_tensor}
\end{equation}
Note that the pressure gradient term in Eq.~(\ref{eq:momentum}) contains the hydrodynamic pressure~$p^{\left(2\right)}$, which is in fact used to enforce continuity instead of solving Eq.~(\ref{eq:continuity}) discretely. The notation is chosen to be consistent with the asymptotic analysis by M{\"u}ller~\cite{Mueller98_lowMach}. \par
Efficient solution of multi-species transport necessitates a simplified description of mass diffusion. In this work, the Curtiss-Hirschfelder approximation~\cite{Hirschfelder54_diff_model} is used with additional consideration of the Soret effect (thermodiffusion). Then, the balance equation for mass fraction $Y_k$ of species~$k$ reads
\begin{align}
\frac{\partial \left( \rho Y_k \right)}{\partial t} + \frac{\partial \left( \rho u_j Y_k \right)}{\partial x_j}  = 
- \frac{\partial}{\partial x_j}\left(  \rho Y_k v_{j,k}  \right)
+ \dot{\omega}_k,
\label{eq:species}
\end{align}
where the effects of multi-species diffusion and thermodiffusion in have been combined into the diffusion velocity 
\begin{equation}
	v_{j,k} = - \left(\textcolor{black}{\frac{D_{\mathrm{th}}}{X_k}} \frac{\partial X_k}{\partial x_j} + \frac{\textcolor{black}{D_{T,k}}}{\textcolor{black}{\rho Y_k} T} \frac{\partial T}{\partial x_j} \right) + v_j^{\mathrm{mc}},
\label{eq:v_diff}
\end{equation}
and~$v_j^{\mathrm{mc}}$ is determined from the condition 
\begin{equation}
\sum_{n=1}^{N_{\mathrm{sp}}}{\rho Y_n v_{j,n}} = 0.
\label{eq:v_corr_cond}
\end{equation}
Note that Einstein's summation convention does not apply to the species index. For simplicity, all species Lewis numbers are set to unity, i.e.\ all diffusivities are equal to the mixture thermal diffusivity ${D_{\mathrm{th}}=\lambda / (\rho c_{p})}$. \textcolor{black}{In Eq.~(\ref{eq:v_diff}), $X_k$  and~$D_{T,k}$ denote the molar fraction and the thermal diffusion coefficient of species~$k$, respectively.}
Chemical source terms in Eq.~(\ref{eq:species}) are computed from a modified kinetic scheme based on the skeletal iso-octane mechanism by Pitsch and Peters~\cite{Pitsch96_iso_octane_mech}, which consists of 26~species and 40~reactions. Reaction rates have been calibrated using the methods by Cai and Pitsch~\cite{Cai14_mech_opt} to match experimental data at elevated pressures~\cite{Galmiche12_sL_exp}.\par
To allow for density calculation from an equation of state, a transport equation for temperature is solved:
\begin{align}
c_{p} \frac{\partial \left( \rho T \right)}{\partial t} + c_{p} \frac{\partial \left( \rho u_j T \right)}{\partial x_j} & = 
\frac{\partial p^{\left(0 \right)}}{\partial t}
+	\frac{\partial}{\partial x_j} \left(\lambda \frac{\partial T}{\partial x_j} \right) \nonumber \\
& -  \rho \frac{\partial T}{\partial x_j} \left[c_{p} v_j^{\mathrm{mc}}  + \sum_{n=1}^{N_{\mathrm{sp}}} c_{p,n}Y_nv_{j,n} \right] \nonumber \\
& - \sum_{n=1}^{N_{\mathrm{sp}}} h_n\dot{\omega}_n + \dot{\mathrm{S}}_{\mathrm{ign}},
\label{eq:temperature}
\end{align}
where the thermal conductivity is denoted as~$\lambda$, the mixture specific heat capacity at constant pressure is named $c_p$, the total specific enthalpy of species $k$ is referred to as $h_k$, and a source term to model spark ignition is introduced as $\dot{\mathrm{S}}_{\mathrm{ign}}$ (cf.\ Sect.~\ref{ssec:ign}). Note that radiation and viscous heating have been neglected. In preliminary investigations using an optically thin gas model, it was found that radiation would reduce the unstretched laminar burning velocity by 0.03\,\% under the present conditions. \par
The local cell density is computed from the ideal gas law
\begin{equation}
\rho = \frac{p^{\left(0\right)}W}{R T},
\label{eq:eos}
\end{equation}
where $R$ is the universal gas constant \textcolor{black}{and $W$ denotes the molecular weight of the mixture}. An additional constraint for constant-volume combustion is given by global mass conservation:
\begin{equation}
m_{\Omega} = \int_{\Omega}\rho \diff V = \mathrm{const.},
\label{eq:glob_mass_cons}
\end{equation}
where the volume integral is computed over the flow domain~$\Omega$. Eqs.~(\ref{eq:eos}) and~(\ref{eq:glob_mass_cons}) are used to update both background pressure~$p^{\left(0\right)}$ and density~$\rho$. With this, the system of equations required to solve the reacting flow problem is closed. \par
To analyze the global heat release rate, a reaction progress variable will be used in Sect.~\ref{sec:react_flow}. If the progress variable definition based on a reduced sum of species mass fractions~\cite{Vervisch00_combust_modeling} was applied in this study, a more complex progress variable transport equation than Eq.~(\ref{eq:math_form_global_srcPROG}) would emerge. This is due to the model expression used to describe multi-species diffusion and due to the Soret effect (cf.\ Eq.~(\ref{eq:species})). To simplify the analysis and make the dataset more accessible for modeling, a new reaction progress variable~$\zeta$ is defined by the solution of the transport equation
\begin{equation}
\frac{\partial \left( \rho \zeta \right)}{\partial t} + \frac{\partial \left( \rho u_j \zeta \right)}{\partial x_j} =
	\frac{\partial}{\partial x_j} \left( \rho D_{\mathrm{th}} \frac{\partial \zeta }{\partial x_j} \right) + \dot{\omega}_{\zeta}.
\label{eq:c0_eq}
\end{equation}
The chemical source term is chosen as the sum of source terms of main product species:
\begin{equation}
\dot{\omega}_{\zeta} = \dot{\omega}_{\mathrm{H}_2}+\dot{\omega}_{\mathrm{H}_2\mathrm{O}} + \dot{\omega}_{\mathrm{CO}} + \dot{\omega}_{\mathrm{CO}_2}.
\label{eq:c0_eq_src_term}
\end{equation}
Note that Eq.~(\ref{eq:c0_eq}) can be obtained from the sum of the transport equations of the major species~(cf.\ Eq.~(\ref{eq:species})), if the Soret effect is neglected and constant molecular weight of the mixture is assumed.
\subsection{Numerical Discretization}
\label{ssec:ciao_discr}
The present DNS database has been generated with the CIAO code~\cite{highq15_ciao} using a time- and space-staggered finite-difference reactive Navier-Stokes solver on Cartesian grids. Mass, momentum, and kinetic energy are conserved discretely in the low-Mach number formulation~\cite{desjardins08_numerics}. In this study, the momentum equations are discretized with a fourth-order-accurate central difference scheme. For the scalar transport equations, a weighted essentially non-oscillatory~(WENO5) scheme~\cite{Jiang96_weno_scheme} is used. This scheme has fourth-order accuracy in smooth monotone regions and third-order accuracy near discontinuities. Using combinations of higher-order finite difference and upwind-biased schemes may result in inconsistencies between discrete mass conservation and scalar convection. To avoid this issue, a finite-volume mass flux is used in the scalar transport equations~\cite{trisjono16_fv}. \par
Time integration is performed with a second-order-accurate Crank-Nicolson scheme~\cite{Crank_Nicolson47}. For time-advancement of the Navier-Stokes equations, a variant of the fractional step method is used~\cite{Kim85_fract_step,Pierce01_diss}. The pressure correction is obtained by solving the Poisson equation with the algebraic multi-grid solver Hypre Boomer AMG~\cite{Falgout02_hypre,Henson02_hypre_boomer_amg}. To efficiently advance the stiff advection-diffusion-reaction equations for species and temperature, the symmetric operator splitting proposed by Strang~\cite{Strang68_splitting} is used. The resulting system of ordinary differential equations for the zero-dimensional homogeneous reactor in each grid cell is solved using a fully time-implicit backward difference method~\cite{Hindmarsh05_sundials,Brown89_cvode}. The Jacobian is evaluated analytically for computational efficiency. \textcolor{black}{All datasets have been computed on grids with~$960^3$ cells}. Numerical accuracy tests performed on one-dimensional laminar flames under DNS conditions are reported in Sect.~1.1 of the supplementary material.

\subsection{Physical Configuration}
\label{ssec:phys_config}
Mixture state and composition have been chosen to be representative for fully homogeneous, stoichiometric SI engine operation at low load. \textcolor{black}{To enable a better understanding of how the flow alters the young flame kernel, Lewis numbers~(Le) of all species have been artificially set to unity. Flame response in presence of differential diffusion \textcolor{black}{is discussed in two separate studies~\cite{Falkenstein19_kernel_Le_I_cnf,Falkenstein19_kernel_Le_II_cnf}}.} Mixture properties of the DNS are summarized in Tab.~\ref{tab:dns_mixture}. \textcolor{black}{The thermodynamic state is given by the pressure~$p^{\left(0\right)} $ and temperature of the unburned gas~$T_{\mathrm{u}}$, which has unity equivalence ratio~$\phi$}. \par
\begin{table}
\caption{Flame conditions in the DNS.}
\vspace{0.1cm}
\centering
\begin{tabular}{P{.35\textwidth}|P{.35\textwidth}} 
\hline
~ & ~ \\ [-10pt]
Property & Value  \\
~ & ~ \\ [-10pt]
\hline
~ & ~ \\ [-10pt]
Mixture   & Iso-Octane/Air \\ 
$p^{\left(0\right)} $ & $6$ bar  \\
$T_{\mathrm{u}}$ & $600$ K \\
$\phi$ & $1.0$ (homogeneous)\\
$\mathrm{Le}$ & $1.0$ \\
Flow Field & Decaying h.i.t. \\
Combust. Regime & Thin Rct. Zones\\
\hline
\end{tabular}
\label{tab:dns_mixture}
\end{table} 
%
A large number of preliminary parameter investigations was performed to match the DNS setup as closely as possible to typical engine conditions reported in the literature. In order to provide well-defined, but engine-relevant flow configurations, all flames have been computed in decaying homogeneous isotropic turbulence. A comparison between engine and DNS parameters is summarized in Tab.~\ref{tab:dns_params}. 
The turbulent integral length scale and eddy turnover time are denoted by~$l_{\mathrm{t}}$ and~$\tau_{\mathrm{t}}$, $\eta$ is the Kolmogorov length scale, while the laminar flame thickness \textcolor{black}{and laminar flame time are referred to as~$l_{\mathrm{f}}$ and~$t_{\mathrm{f}}$, respectively.} 
In this study, the laminar flame thickness~$l_{\mathrm{f}}$ is computed from the maximum temperature gradient, while a turbulent length scale representative for the integral scales is given by $l_{\mathrm{t}} = {{u^{\prime}}^3}/{\overline{\varepsilon}}$, using the mean dissipation rate $\overline{\varepsilon}$. 
Except for the integral length scale of turbulence, which is approximately~2.5 times smaller than in a practical engine, the conditions are realistic for the full duration of the DNS. More details on the flow field initialization and non-reactive DNS results are provided as supplementary material. \par
\begin{figure}
\centering
\begin{minipage}[b]{0.225\textwidth} 
  \centering
\makebox[0pt][l]{~}
\end{minipage}
\begin{minipage}[b]{0.45\textwidth} 
  \centering
 \includegraphics[angle=0,origin=c,width=0.95\textwidth]{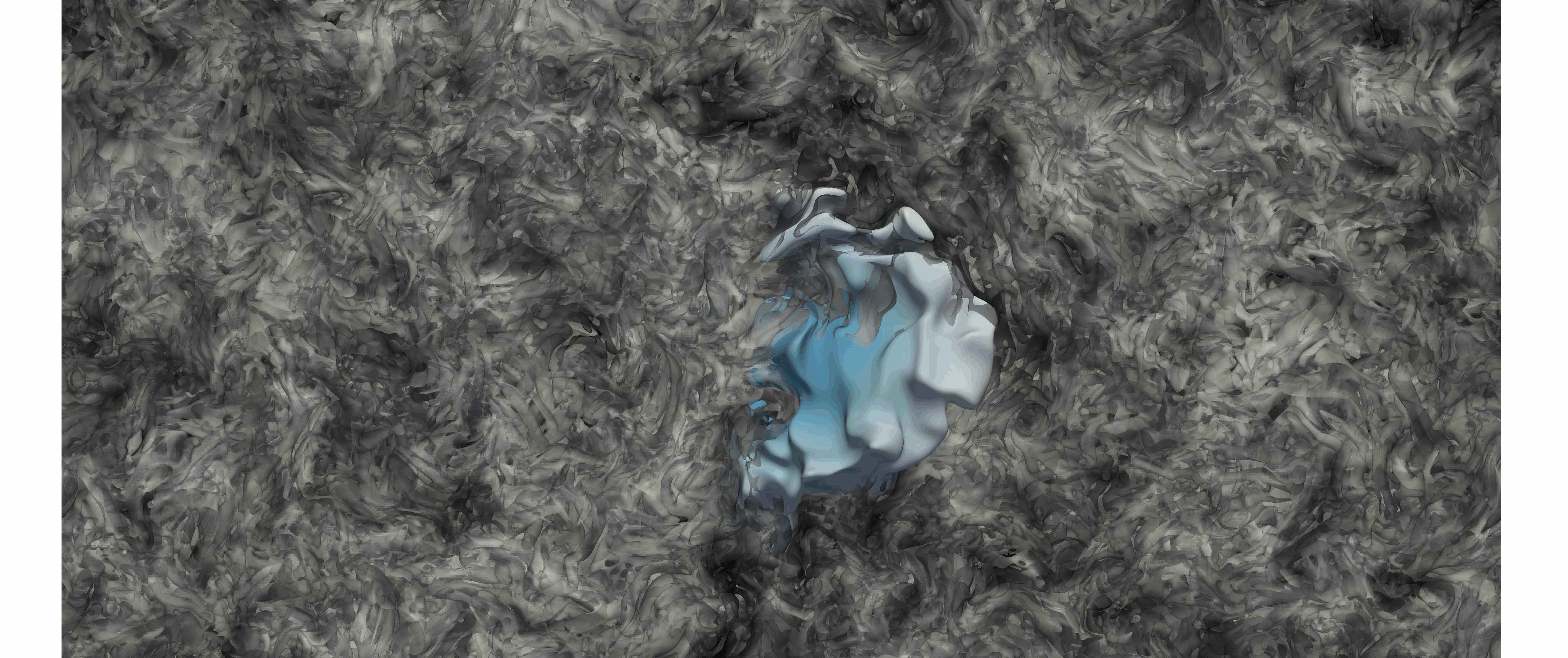}
\end{minipage}
\begin{minipage}[b]{0.225\textwidth} 
  \centering
\makebox[0pt][l]{~}
\end{minipage}
\begin{minipage}[b]{0.225\textwidth} 
  \centering
\makebox[0pt][l]{~}
\end{minipage}
\begin{minipage}[b]{0.45\textwidth}
  \centering
 \includegraphics[angle=0,origin=c,trim={-7cm 0cm -8cm 0cm},clip,width=0.95\textwidth]{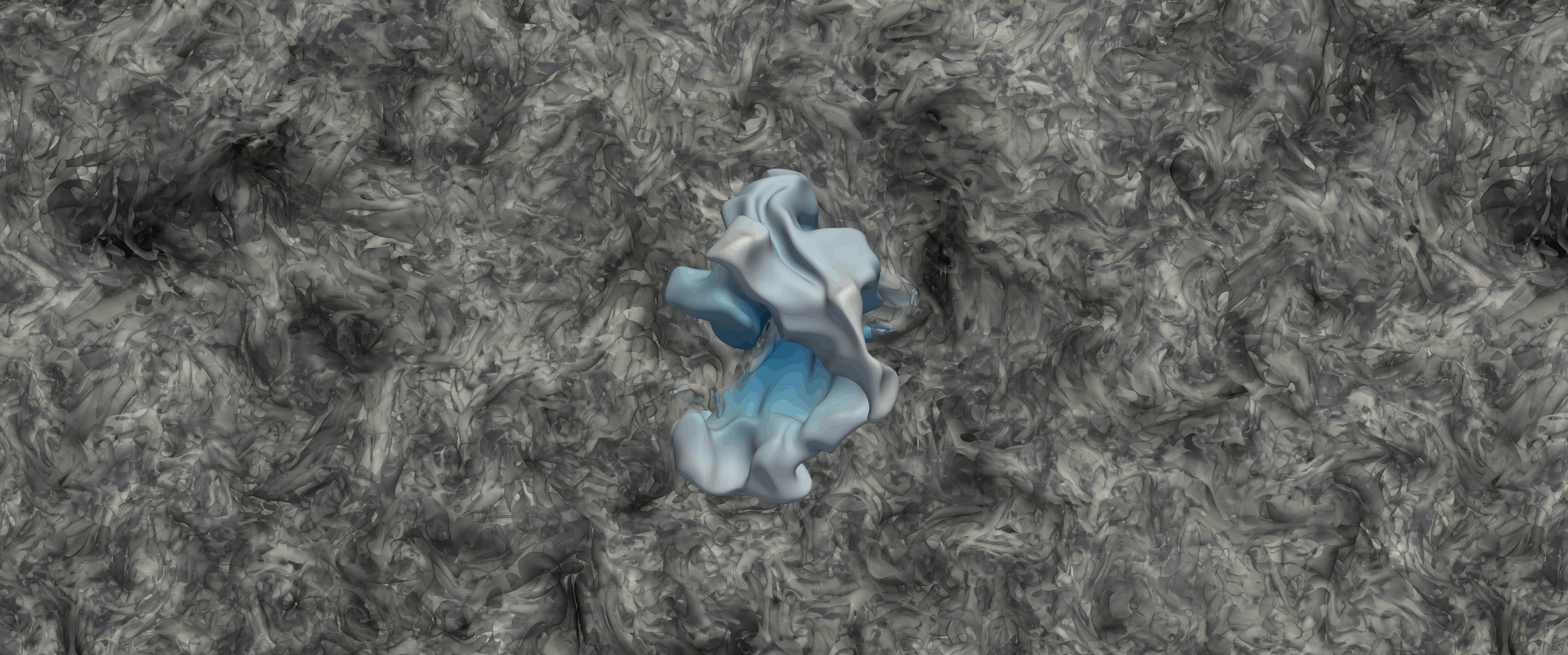}
\end{minipage}
\begin{minipage}[b]{0.225\textwidth} 
  \centering
\makebox[0pt][l]{~}
\end{minipage}
\begin{minipage}[b]{0.225\textwidth} 
  \centering
\makebox[0pt][l]{~}
\end{minipage}
\begin{minipage}[b]{0.45\textwidth}
  \centering
   \includegraphics[angle=0,origin=c,width=0.95\textwidth]{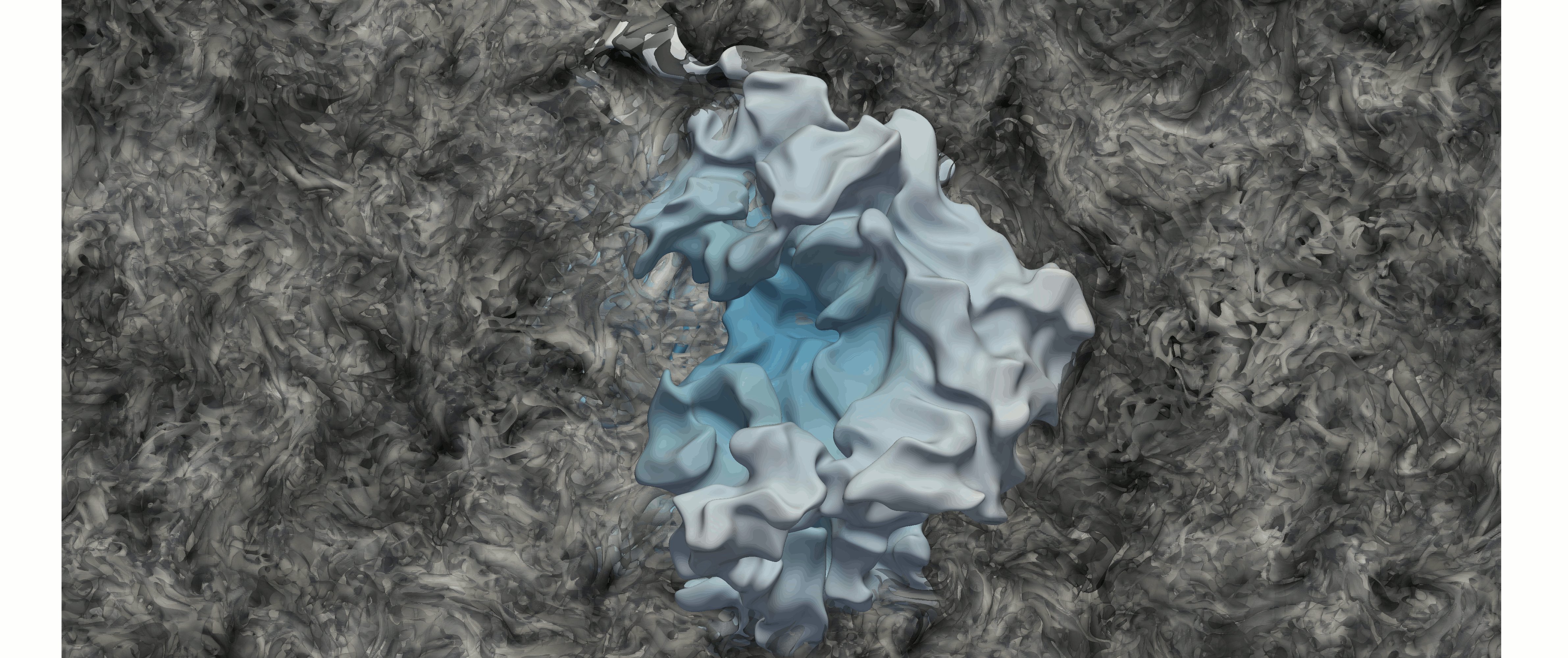}
\end{minipage}
\begin{minipage}[b]{0.225\textwidth} 
  \centering
\makebox[0pt][l]{~}
\end{minipage}
\begin{minipage}[b]{0.225\textwidth} 
  \centering
\makebox[0pt][l]{~}
\end{minipage}
\begin{minipage}[b]{0.45\textwidth}
  \centering
     \includegraphics[angle=0,origin=c,width=0.95\textwidth]{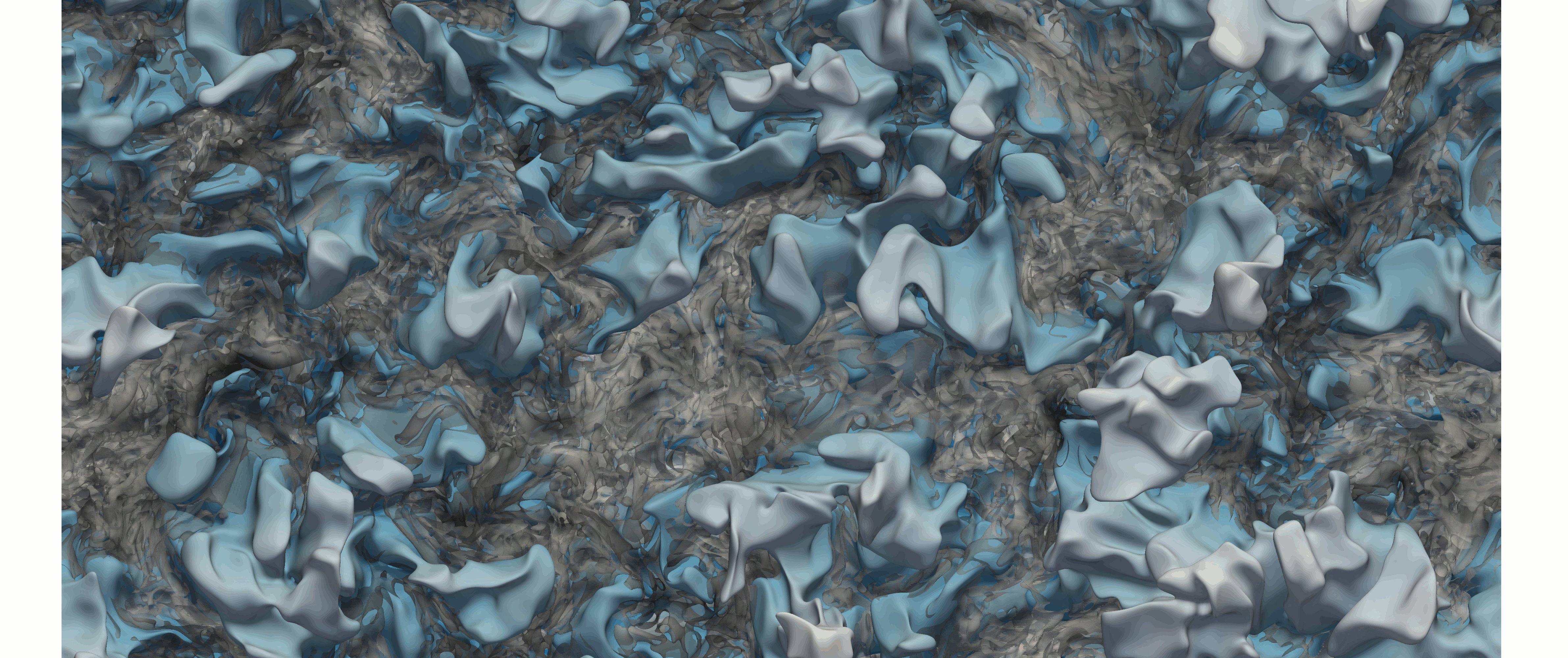}
\end{minipage}
\begin{minipage}[b]{0.225\textwidth} 
  \centering
\makebox[0pt][l]{~}
\end{minipage}
\caption{Iso-surfaces of temperature (with arbitrary color scale) and the second invariant of the velocity gradient tensor (grey background) for the four flames at $t=1.0\cdot\tau_{\mathrm{t}}$.}
\label{fig:flame_geom_vis}
\end{figure}

\begin{table} 
\caption{Engine \citep{Heywood94_COMODIA,Heim11_engine_turb} and DNS characteristic numbers.}
\vspace{0.1cm}
\centering
\begin{tabular}{P{.23\textwidth}|P{.23\textwidth}|P{.23\textwidth}}  
\hline
~ & ~ & ~ \\ [-10pt]
 Parameter   & Engine  & DNS ($t_{\mathrm{init}} - t_{\mathrm{end}}$) \\
~ & ~ & ~ \\ [-10pt]
\hline
~ & ~ & ~ \\ [-10pt]
    $\mathrm{Re}_{\mathrm{t}} $& $100-2390$  & $ 385 - 222$\\[3pt]
	$\frac{u_{\mathrm{rms}}}{\mathrm{s}_{\mathrm{l}}^{\sm{0}}}$& $2-15$ & $ 7.0 - 3.1 $\\[3pt]
	$\mathrm{Ka} $& $1-6$ & $13.7 - 4.2$\\[3pt]
	$\mathrm{Da}$ &$1-100$ & $ 1.4 - 3.6$\\[3pt]
	$\frac{l_{\mathrm{t}}}{\eta}$& 100 -200 & $87.2 - 57.2$\\[3pt]  
	$\frac{l_{\mathrm{t}}}{l_{\mathrm{f}}}$& $20 - 147$ & $10.2 - 12.4$\\[3pt]
	\hline
~ & ~ & ~ \\ [-10pt]
		$\frac{D_{0}}{l_{\mathrm{t}}}$ &$<1.0$ & $ 0.3 \;|\; 2.0 \; |\; \infty $\\[3pt]
	\hline
  \end{tabular}
  \label{tab:dns_params}
\end{table} 
\textcolor{black}{In this work, we seek to gain quantitative insight into early flame kernel development under engine conditions by means of three-dimensional DNS with detailed chemistry. Due to high computational cost, an extensive parameter variation is out of scope. Instead, only the ratio of initial flame diameter~$D_{0}$ to the integral length scale~$l_{\mathrm{t}}$ is varied, while the ratio of velocity fluctuation to burning velocity is kept constant. \textcolor{black}{$D_{0}$ is determined as the equivalent diameter of the burned volume at the time when the ignition heat source is switched off.} The following datasets have been generated and will be analyzed in detail:
\begin{itemize}
\item{Engine flame kernel, $D_{0}/l_{\mathrm{t}}=0.3$,}
   \begin{itemize}
     \item Engine Kernel I: $t_{\mathrm{sim.}}=2.3\cdot\tau_{\mathrm{t}}$, \\
                            \makebox[88pt][l]{}\textcolor{black}{$x_{\mathrm{ign.}} = (0,0,0)$.}
     \item Engine Kernel II: $t_{\mathrm{sim.}}=1.0\cdot\tau_{\mathrm{t}}$, \\
                            \makebox[93pt][l]{}\textcolor{black}{$x_{\mathrm{ign.}} = (5\,l_{\mathrm{t}},5\,l_{\mathrm{t}},5\,l_{\mathrm{t}})$.}
   \end{itemize}
\item{Large flame kernel, $D_{0}/l_{\mathrm{t}}=2.0$,}
\item{Planar flame, $D_{0}/l_{\mathrm{t}}=\infty$.}
\end{itemize}
\textcolor{black}{Here, the ignition location is specified with respect to the center of the computational domain}. While flame kernels were computed in a constant volume setting, an outflow boundary condition ahead of the planar flame was used to avoid excessive pressure increase. \textcolor{black}{For all simulations, a domain size of $15 \, l_{\mathrm{t}}$ was used.} The flow field in all three flame configurations is very similar as shown by the mean strain rate conditioned on the distance to the flame front, which is depicted in Fig.~\mbox{S-7} in the supplementary material. Visualizations of all flames at approximately one eddy turnover time after flame initialization are given in Fig.~\ref{fig:flame_geom_vis}. Additional illustrations of small-flame kernel growth are provided in Figs.~\mbox{S-12} and~\mbox{S-13} in the supplementary material.} \par
\begin{figure}
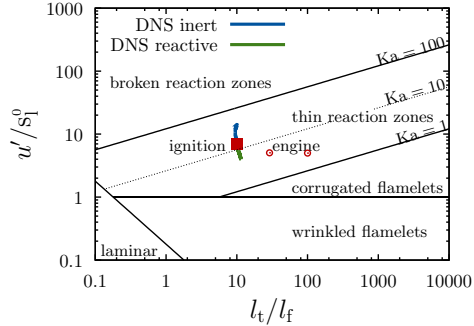

\centering
\input{template/change_font_10.tex} 
\scalebox{0.7}{\input{./data/FLAME_KERNEL_DNS_02/regime_diagram/dns_regime_diagram_TOBIAS_JFM_Ka10_HALF.tex}}
\input{template/change_font_12.tex} 
\caption{History of DNS flame conditions in the regime diagram of premixed turbulent combustion. In addition, two typical engine operating points for low (left red point) and high engine load (right red point) are shown.}
\label{fig-prem_comb_regime_diagr}
\end{figure}

\graphicspath{{./figures/animation/}}
Flame conditions throughout the DNS are illustrated in the regime diagram of turbulent premixed combustion~\cite{Peters99_regime_diagr} in Fig.~\ref{fig-prem_comb_regime_diagr}. The diagram has been adjusted to be representative for the unburned mixture defined in Tab.~\ref{tab:dns_mixture}. More details on the definitions of the characteristic numbers in the diagram are discussed in Sect.~1.3 of the supplementary material.
In Fig.~\ref{fig-prem_comb_regime_diagr}, it can be seen that the present DNS is parameterized such that it falls into the thin reaction zones regime, which is commonly the case for SI engine combustion. For reference, two operating points representing low (left point) and high engine load (right point) are marked in the figure. While the laminar burning velocity is almost constant during the DNS, turbulent kinetic energy decays over time. Consequently, the velocity ratio ${u^{\prime}}/{\mathrm{s}_{\mathrm{l}}^{\sm{0}}}$ decreases towards the end of the simulation.
\graphicspath{{./data/FLAME_KERNEL_DNS_02/regime_diagram/}}
\subsection{Flame Initialization}
\label{ssec:ign}
To be consistent with realistic engine flame development, an ignition heat source has been employed to initialize the flame kernels. The source term is defined to vary smoothly in time and space such that the low-Mach number flow and ideal gas regimes remain valid. A non-dimensional time is introduced using the ignition duration $\tau_{\mathrm{ign}}$ as
\begin{equation}
\hat{t}_{\mathrm{ign}} = \frac{t-t_{\mathrm{ign}}}{\tau_{\mathrm{ign}}}.
\label{eq:ign_t_norm}
\end{equation}
The ignition source term in Eq.~(\ref{eq:temperature}) is given by
\begin{equation}
 \dot{\mathrm{S}}_{\mathrm{ign}}\left(\hat{t}_{\mathrm{ign}},r\right) = \left\{
    \begin{array}{ll}
	\frac{\psi\left(\hat{t}_{\mathrm{ign}} \right)}{\int_0^1 \psi \diff\hat{t}}      
      \frac{\mathrm{E}_{\mathrm{ign}}}{c_{p} \tau_{\mathrm{ign}} V_{\mathrm{ign}}} 
\left[1 - \exp\left(- \left[\frac{\pi}{2} - \frac{r}{R_{\mathrm{ign}}} \right]^4\right) \right], & r\leq \frac{\pi}{2} R_{\mathrm{ign}}  \\[2pt]
      0, & \mathrm{otherwise},
  \end{array} \right.
  \label{eq:ign_Sdot}
\end{equation}
where a spherical reference volume is defined by ${V_{\mathrm{ign}}=4/3\pi R^3_{\mathrm{ign}}}$ for a given ignition radius $R_{\mathrm{ign}}$, and $\mathrm{E}_{\mathrm{ign}}$ denotes an ignition energy parameter. The time-dependent scaling factor $\psi\left(\hat{t} \right)$ is determined by
\begin{equation}
\psi\left(\hat{t} \right) = \frac{2 \cdot \hat{t}^2}{\mathrm{C}^2}
\frac{\log\left(\hat{t}\right)}{\log\left(\mathrm{C}\right)},
\label{eq:ign_psi_scale}
\end{equation}
where $\mathrm{C}=0.603$ is a constant. For the engine flame kernel configuration, the ignition parameters were selected as $\tau_{\mathrm{ign}} \textcolor{black}{= 0.16\,t_{\mathrm{f}}} = 15\,\mu\mathrm{s} $, $R_{\mathrm{ign}}=4\,l_{\mathrm{f}}=0.27\,\mathrm{mm}$ and $\mathrm{E}_{\mathrm{ign}}=0.32\,\mathrm{mJ}$ to rapidly initiate a flame, but avoid the occurrence of excessive temperatures. \textcolor{black}{These parameter choices are consistent with previous studies, where similar spark duration~\cite{Hesse09_spark_duration} and ignition kernel size~\cite{Subramanian09_spark_src,Uranakara17_ign_kernel_3d} settings were used. From experimental investigations it is known that in absence of excessive heat losses to the spark plug electrodes, ignition is generally more difficult to achieve in turbulent flows as compared to quiescent mixtures~\cite{Shy17_turb_facilitated_ign}. In the present work, a 40\,\% higher ignition energy than the minimum ignition energy~(MIE) of a laminar, non-unity-Lewis-number flame was found to reliably initiate a flame kernel in multiple realizations of the initial turbulent flow field. The resulting excess temperature in the burned region of the flame kernels mostly decays until ($t\approx1.0\cdot\tau_{\mathrm{t}}$), as shown in Fig.~\mbox{S-6} of the supplementary material.} \par
In case of the planar flame, applying a heat source for flame initialization would activate a large volume expanding in only one spatial direction, which would lead to significant artificial perturbations of the surrounding flow field. Therefore, the planar flame has been introduced into the simulation domain by mapping a one-dimensional laminar unstretched flame solution onto the DNS grid. 
\subsection{Engine Relevance of the DNS Setup}
\label{sec:engine_relevance}
In this section, the practical relevance of the small-flame-kernel DNS datasets 
will be discussed, which are the main reference flames discussed in this study. To make DNS of early flame kernel development under engine conditions computationally feasible, several assumptions and simplifications have been introduced. First, a periodic simulation domain has been used, which acts as a constant-volume combustion chamber. In a real engine, the in-cylinder volume changes due to piston motion. However, since early flame kernel development takes place close to top dead center~(TDC), piston displacement and associated volumetric change are rather small. Due to the small kernel size compared to the total in-cylinder volume, isobaric combustion can be assumed. In the DNS, expansion of the burned gas induces an increase in background pressure and temperature of the unburned mixture by~8.5 and~2\,\%, respectively. From one-dimensional simulations it was found that compared to initial conditions, the laminar burning velocity is increased by~2\,\%, while the laminar flame thickness is reduced by~7\,\% at the end of the DNS run time. \par
Second, the flame kernel develops in decaying turbulence. In fact, experimental investigations in model engines or compression machines have shown that the assumption of decaying turbulence in the late compression or early expansion stroke is justified, even in presence of a coherent vortex structure which breaks down as the piston approaches top dead center~\cite{Boree02_vortex_breakdown,Voisine11_vortex_breakdown}. \par
Third, solid walls, particularly the presence of a spark plug, have been neglected. Under certain in-cylinder flow conditions, the flame kernel may stay attached to the spark plug which then acts as a flame holder. Associated heat losses and flow/wall interactions will likely lead to very different flame kernel behavior than observed in this study. \textcolor{black}{When stochastic flame kernel detachment from the spark plug occurs in an engine, the associated variations in heat loss were found to cause significant cycle-to-cycle variations in early flame kernel growth~\cite{Pischinger91_heat_loss}.} 
It is considered that the present DNS dataset represents situations where the flame kernel is subject to convection and detaches from the spark plug, as observed in experiments reported by Dahms et al.~\cite{Dahms11_spark_exp}.  \par
Fourth, the integral length scale is two to three times smaller in the DNS than real engines. Still, the present turbulent flame conditions fall into the thin reaction zones regime, which is typical for SI engines. With respect to early flame kernel/flow interactions, regime diagrams proposed in the literature distinguish different regimes based on the kernel size relative to a flow length scale, and a flow velocity scale relative to a characteristic burning velocity~\cite{Echekki07_kernel_vortex_map_01,Vasudeo10_kernel_vortex_map_01,Reddy11_regime_map}. Although not intrinsic to these diagrams, the Damk{\"o}hler number can be expected to be an important parameter in flame kernel/flow interactions~\cite{Giannakopoulos18_kernel_dns_3d}. In the present dataset, the turbulence intensity was chosen to be at the lower end of engine-relevant conditions. Hence, a realistic Damk{\"o}hler number could be achieved~\cite{Abraham85_engine_Da}.  \par
In addition to these necessary model abstractions, several simplifications have been introduced to first clarify the impact of turbulence on the flame and allow for systematic future studies. Of primary interest will be \textcolor{black}{flame kernel/wall interactions}, influences of scalar inhomogeneities, as well as flame kernel development in non-equilibrium turbulence. \par
\section{Analysis of the Global Heat Release Rate}
\label{sec:react_flow}
The following analysis aims at clarifying the effects of turbulence on the integral heat release rate during flame kernel development, which according to Eq.~(\ref{eq:key_params}) is influenced by two quantities, the averaged local flame displacement speed~$\mathrm{s}_{\mathrm{d}}$ and the total flame surface area~$\overline{A}_{\Omega}$ \textcolor{black}{or global flame surface density~$\overline{\Sigma}_{\Omega}$}. Hence, the temporal evolution of these governing quantities will be investigated \textcolor{black}{and run-to-run variations will be analyzed}. \par
\begin{figure}
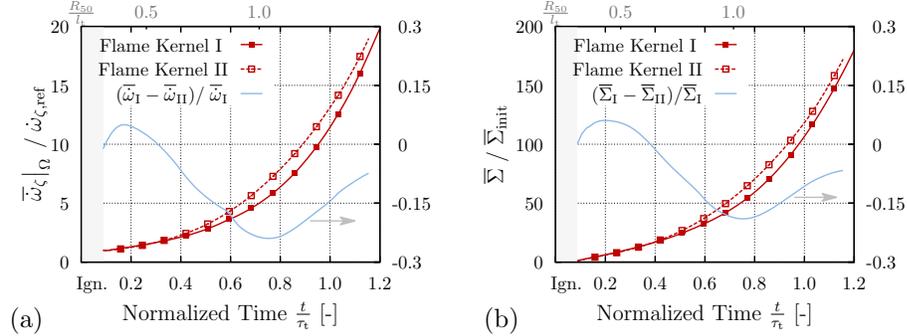

\centering
\begin{minipage}[b]{0.45\textwidth}
  \centering
\input{template/change_font_10.tex} 
  \makebox[0pt][l]{\quad(a)}\scalebox{0.7}{\input{./data/FLAME_KERNEL_DNS_02/181003_curv_strain_kin_restor_diss_divg_skew/kernel_dns_mean_srcPROG_time_01_02_JFM_HALF.tex}}
\input{template/change_font_12.tex} 
\end{minipage}
\begin{minipage}[b]{0.45\textwidth}
  \centering
\input{template/change_font_10.tex} 
   \makebox[0pt][l]{\quad(b)}\scalebox{0.7}{\input{./data/FLAME_KERNEL_DNS_02/181003_curv_strain_kin_restor_diss_divg_skew/flame_kernel_gen_fsd_size_K02_LOC01_JFM_HALF.tex}}
\input{template/change_font_12.tex} 
\end{minipage}
\caption{\textcolor{black}{Integrated progress variable source term~(a) and global flame surface density~(b) during early flame kernel development}.}
\label{fig:kernel_02_srcProg_Sigma_I0}
\end{figure}

\textcolor{black}{The flame-integrated progress variable source term~$\left.\overline{\dot{\omega}_{\zeta}}\right|_{\Omega}$ has been computed for both engine-relevant kernel realizations and is plotted in Fig.~\ref{fig:kernel_02_srcProg_Sigma_I0}\,(a) as function of time. Although the present stoichiometric flame conditions with fully homogeneous unburned mixture and without any heat losses or flame/wall interactions represent rather stable engine combustion, the heat release rate of Engine Kernel~I is temporarily up to 25\,\% lower than in Engine Kernel~II. While the difference in stretch factor~$\overline{\mathrm{I}}_0$ between the flame kernel realizations is less than 5.5\,\% in the present unity-Lewis-number limit, the global flame surface density~$\overline{\Sigma}_{\Omega}$ evolution exhibts stronger run-to-run variations (cf.\ Fig.~\ref{fig:kernel_02_srcProg_Sigma_I0}\,(b)). This observation is somewhat expected since flame surface area production is governed by the low-wavenumber range of the turbulent energy spectrum, i.e.\ flame kernels of comparable size to the integral length scale are not yet exposed to a statistically homogeneous flow field. However, the flame structure and displacement speed might be altered by small-scale turbulent mixing, which will be investigated in the next section. A detailed analysis of flame area dynamics will be presented in Sect.~\ref{ssec:flame_area}.}
\graphicspath{{./data/FLAME_KERNEL_DNS_02/181003_curv_strain_kin_restor_diss_divg_skew/}}
%

\subsection{Mean Flame Displacement Speed Evolution}
\label{ssec:mean_Sd}
In order to quantify deviations in displacement speed from a laminar unstretched premixed flame, the stretch factor~$\mathrm{I}_0$ (cf.\ Eq.~(\ref{eq:I0_def_01})) is considered. \textcolor{black}{In addition to stretch effects,} $\mathrm{I}_0$ quantifies the acceleration of the flame by ignition, and the amplification of molecular mixing by small-scale turbulence according to Damk{\"o}hler's second hypothesis~\cite{Damkoehler40}. Since flame conditions in SI engines \textcolor{black}{operated with stoichiometric mixture and without exhaust gas recirculation~(EGR)} are typically situated close to the unity-Karlovitz-number boundary of the thin reaction zones regime, i.e.\ Karlovitz numbers range between~1 and~10 (cf.\ Fig.~\ref{fig-prem_comb_regime_diagr}), an assessment of deviations from the corrugated flamelet behavior is desirable in terms of modeling. \par
For a comprehensible analysis, the flame-integrated stretch factor~$\overline{\mathrm{I}}_0$ is split into components by normal-propagation and by tangential diffusion~(cf.\ Eq.~\ref{eq:sd_thin_rct_zones}):
\begin{align}
\overline{\mathrm{I}}_0^{(\zeta)} & = \frac{\left< \rho {\mathrm{s}}^{(\zeta)}_{\mathrm{d}} \right>_{\mathrm{s,}\Omega}}{\rho_{\mathrm{u}}\mathrm{s}_{\mathrm{l,u}}^{\sm{0}}}
= \frac{\left< \rho \left( {\mathrm{s}}^{(\zeta)}_{\mathrm{rn}} + {\mathrm{s}}^{(\zeta)}_{\kappa} \right) \right>_{\mathrm{s,}\Omega}}{\rho_{\mathrm{u}}\mathrm{s}_{\mathrm{l,u}}^{\sm{0}}} \nonumber\\
& =  \overline{\mathrm{I}}^{(\zeta)}_{0,\mathrm{rn}} + \overline{\mathrm{I}}^{(\zeta)}_{0,\kappa},
\label{eq:I0_def_split}
\end{align}
where
\begin{align}
\overline{\mathrm{I}}^{(\zeta)}_{0,\mathrm{rn}} &= \frac{\left< \rho {\mathrm{s}}^{(\zeta)}_{\mathrm{rn}}  \right>_{\mathrm{s,}\Omega}}{\rho_{\mathrm{u}}\mathrm{s}_{\mathrm{l,u}}^{\sm{0}}}  ,\label{eq:I0_def_split_rn} \\
\overline{\mathrm{I}}^{(\zeta)}_{0,\kappa} & = \frac{\left< \rho {\mathrm{s}}^{(\zeta)}_{\kappa} \right>_{\mathrm{s,}\Omega}}{\rho_{\mathrm{u}}\mathrm{s}_{\mathrm{l,u}}^{\sm{0}}} 
= \frac{\left< -\rho D_{\mathrm{th}} \kappa^{(\zeta)} \right>_{\mathrm{s,}\Omega}}{\rho_{\mathrm{u}}\mathrm{s}_{\mathrm{l,u}}^{\sm{0}}}. \label{eq:I0_def_split_k}
\end{align}
Here, $\kappa^{(\zeta)} $ denotes the mean curvature computed from the scalar gradient field of the reaction progress variable~$\zeta$, which is defined by the solution of Eq.~(\ref{eq:c0_eq}). Note that this definition simplifies the analysis, since complex diffusion effects, such as thermodiffusion, are implicitly contained in~$\zeta$ and no additional terms other than those given in Eq.~(\ref{eq:I0_def_split}) must be considered when splitting the displacement speed into its components. The numerical formulation derived for conservative splitting of the diffusion term is provided in Sect.~1.6 of the supplementary material. With respect to the sign convention used in this work, normal vectors are directed towards negative scalar gradients:
\begin{equation}
 n^{(\zeta)}_i = - \frac{1}{\left|\nabla \zeta \right|} \frac{\partial \zeta}{\partial x_i}. 
\label{eq:nvec_def}
\end{equation}
For brevity, the superscript indicating the scalar field from which the quantities have been computed will be omitted hereafter. Note that from the definition of the normal vector in Eq.~(\ref{eq:nvec_def}) it follows that surfaces with convex shape towards the unburned gas (such as spherical expanding flames) have positive curvature. \par
 Small changes in unburned mixture thermodynamic state due to constant-volume flame kernel DNS are accounted for by time-varying laminar reference data in the denominators of Eqs.~(\ref{eq:I0_def_split_rn}) and~(\ref{eq:I0_def_split_k}). Global flame quantities are obtained by computing the flame-surface-average over all scalar iso-surfaces (cf.\ Eq.~(\ref{eq:gen_fsd_avg})). \par
\begin{figure}
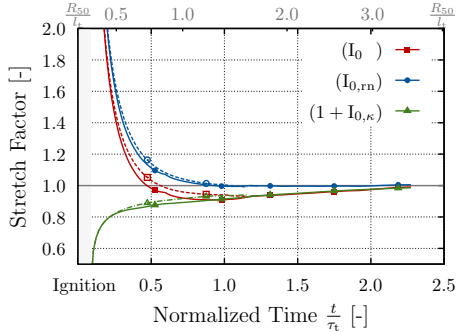

  \centering
\input{template/change_font_10.tex} 
  \scalebox{0.7}{\input{./data/FLAME_KERNEL_DNS_02/181003_curv_strain_kin_restor_diss_divg_skew/kernel_dns_mean_I0_Sd_1d_time_02_LOC01_JFM_HALF.tex}}
\input{template/change_font_12.tex} 
\caption{Flame displacement speed deviation from a laminar unstretched flame \textcolor{black}{for Engine Kernel~I (solid lines) and Engine Kernel~II (dashed lines)}. \textcolor{black}{The upper x-axis indicates the size of Engine Kernel~I.}} 
\label{fig:kernel_02_mean_I0}
\end{figure}

In Fig.~\ref{fig:kernel_02_mean_I0}, $\overline{\mathrm{I}}_0$ as well as the two contributions by normal propagation and tangential diffusion are depicted as function of time for \textcolor{black}{both engine flame kernel realizations, which show very similar behavior. Note that the upper x-axis indicates the size of Engine Kernel~I.} Initially, ignition energy supply causes rapid flame propagation. Due to the small heat source diameter, large mean curvatures induce a noticeable reduction in displacement speed by~$\overline{\mathrm{I}}_{0,\kappa}$. As the flame kernels expand and mean curvature declines, the integral effect of tangential diffusion diminishes.
The time history of~$\overline{\mathrm{I}}_{0,\mathrm{rn}}$ shows that ignition effects decay before one turbulent time scale is reached ($t=0.8\cdot\tau_{\mathrm{t}}$), i.e.\ the normal-propagation component reaches a constant value of unity. 
In other words, the mean normal displacement speed of the turbulent expanding flame is identical to the laminar burning velocity of an unstretched premixed flame. This indicates that local turbulent perturbations of flame structure do on average not lead to enhanced mixing under the considered engine-relevant Karlovitz-number conditions with unity Lewis number. \par
\graphicspath{{./data/FLAME_KERNEL_DNS_02/181003_curv_strain_kin_restor_diss_divg_skew/}}
Still, it is to be shown to which extent small-scale turbulent mixing alters the flame structure in the present dataset and how fast such perturbations develop. For this purpose, temperature mean and standard deviation profiles will be considered as function of spatial coordinate. The origin of the local one-dimensional coordinate system is located at the temperature level corresponding to the maximum heat release rate in a laminar unstretched flame. This choice is motivated by the expected laminar flamelet behavior of the inner layer in the thin reaction zones regime~\cite{Peters99_regime_diagr}. A signed distance function field relative to this reference iso-surface is computed by using a parallel implementation of the fast marching method~\cite{Herrmann03_fmm}. \par
\begin{figure}
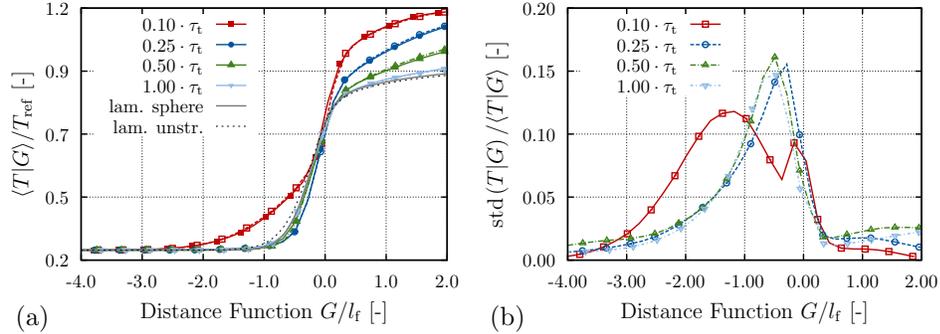

\centering
\begin{minipage}[b]{0.45\textwidth}
  \centering
\input{template/change_font_10.tex} 
  \makebox[0pt][l]{\quad(a)}\scalebox{0.7}{\input{./data/FLAME_KERNEL_DNS_02/180813_dist_fct_strain_rate_T/kernel_dns_Tmean_G_early_02_LOC01_JFM_HALF.tex}}
\input{template/change_font_12.tex} 
\end{minipage}
\begin{minipage}[b]{0.45\textwidth}
  \centering
\input{template/change_font_10.tex} 
  \makebox[0pt][l]{\quad(b)}\scalebox{0.7}{\input{./data/FLAME_KERNEL_DNS_02/180813_dist_fct_strain_rate_T/kernel_dns_Tvar_G_early_02_JFM_HALF.tex}}
\input{template/change_font_12.tex} 
\end{minipage}
\caption{Engine flame kernel: Conditional mean temperature \textcolor{black}{(two kernel realizations)}~(a) and standard deviation of temperature \textcolor{black}{(only one kernel)}~(b) as function of distance to the iso-surface of max.\ heat release rate shortly after ignition.}
\label{fig:kernel_02_Tmean_Tvar_G_early}
\end{figure}

In Fig.~\ref{fig:kernel_02_Tmean_Tvar_G_early}\,(a), conditional mean temperature profiles extracted from the small flame kernel dataset are compared to two laminar reference flames. Besides the common laminar unstretched configuration, an unsteady spherical expanding flame has been computed using parameters analogous to the flame kernel DNS. The plotted temperature profile has been extracted at a time corresponding to ($t=1.7\cdot\tau_{\mathrm{t}}$).
As can be seen from the engine flame kernel data shown in Fig.~\ref{fig:kernel_02_Tmean_Tvar_G_early}\,(a), the flame structure \textcolor{black}{of both realizations} is initially determined by the ignition heat source. At ($t=0.1\cdot\tau_{\mathrm{t}}$), the burned gas region reaches temperatures well above the adiabatic flame temperature, while ahead of the flame, increased temperatures are observed up to a distance of approximately two times the laminar flame thickness. As the young flame expands, ignition artefacts in the preheat zone rapidly disappear. At ($t=0.5\cdot\tau_{\mathrm{t}}$), the low-temperature region of the laminar spherical expanding reference flame is almost recovered by the conditional mean data. Compared to the unstretched reference flame, the preheat region of the spherical flames is slightly thinner, possibly due to tangential diffusion effects. In the high-temperature region, the conditional mean profile of the turbulent flame kernel approaches the laminar reference flame structure after ($t=1.0\cdot\tau_{\mathrm{t}}$), which is consistent with the decay of ignition effects observed in Fig.~\ref{fig:kernel_02_mean_I0}. 
\graphicspath{{./data/FLAME_KERNEL_DNS_02/180813_dist_fct_strain_rate_T/}}
\textcolor{black}{It is shown in Fig.~\mbox{S-9} of the supplementary material that the conditional mean temperature profile remains very similar to a laminar flame structure until the end of the simulation. Hence, the turbulent flame structure is on average not thickened by small-scale turbulent mixing under the present conditions with Karlovitz numbers below~14, as already concluded from the near-laminar normal displacement speed which was discussed in the context of Fig.~\ref{fig:kernel_02_mean_I0}.} \par
To quantify the degree of turbulent mixing inside the flame structure of the small kernel, conditional temperature standard deviation profiles as function of distance to the reference iso-surface are depicted in Fig.~\ref{fig:kernel_02_Tmean_Tvar_G_early}\,(b). Shortly after ignition, locally increased temperatures are present up to four times the laminar flame thickness ahead of the ignition kernel. As expected, due to the short eddy turnover times at small length scales, temperature fluctuations inside the preheat region develop quickly. At ($t=0.25\cdot\tau_{\mathrm{t}}$), the standard deviation reaches up to 15\,\% of the mean, and is very close to the profiles extracted at later times. Since turbulence is decaying in the present configuration, small-scale eddies are dissipated as time proceeds. Consequently, temperature fluctuations slightly decrease in magnitude towards the end of the simulation (cf.\ Fig.~\ref{fig:kernel_02_Tmean_Tvar_G_early},(b) and Fig.~\mbox{S-9}\,(b) in the supp. material).\par
It has been shown that in the engine-relevant flame kernel datasets, small-scale turbulent mixing develops quickly ($t<0.25\cdot\tau_{\mathrm{t}}$), while ignition effects are visible until ($t=0.8\cdot\tau_{\mathrm{t}}$). Interestingly, turbulence does not seem to alter the mean displacement speed of the small flame kernel under the present Karlovitz-number ($4<\mathrm{Ka}<14$), fully premixed engine conditions at unity Lewis number, which are located well inside the thin reaction zones regime. However, there is a chance that the flame structure is preferentially perturbed in regions of reduced scalar gradients, e.g.\ where curvatures are high. 
Results extracted from the large flame kernel and planar flame datasets are not shown for brevity, but lead overall to the same conclusions. \par
\subsection{Evolution of Flame Surface Area}
\label{ssec:flame_area}
\textcolor{black}{As it was shown in Fig.~\ref{fig:kernel_02_srcProg_Sigma_I0}\,(a), significant run-to-run variations in the global heat release rate of both engine-relevant flame kernel realizations exist, which must be reflected in the evolution of flame structure and/or flame geometry according to Eq.~(\ref{eq:key_params}).  While the flame displacement speed was shown to be similar in both kernel realizations (cf.\ Fig.~\ref{fig:kernel_02_mean_I0}), noticeable differences in global flame surface density (total flame area) were observed during the early kernel development phase (cf.\ Fig.~\ref{fig:kernel_02_srcProg_Sigma_I0}\,(b)). }
According to Damk{\"o}hler's first hypothesis~\cite{Damkoehler40}, flame/flow interactions are purely kinematic in the limit of large-scale turbulence, i.e.\ if the smallest turbulent length scales are larger than the laminar flame thickness. This asymptotic case corresponds to the corrugated flamelet regime in Fig.~\ref{fig-prem_comb_regime_diagr}, which is in fact very close to the present engine-relevant conditions, where small-scale mixing exists only to limited extent (cf.\ Sect.~\ref{ssec:mean_Sd}). Since the flame kernel diameter is initially on the order of the integral length scale, early flame front/turbulence interactions cannot be expected to follow statistical behavior. \textcolor{black}{In this section, 
run-to-run variations in net area production observed in the present engine-relevant flame kernel configuration and previous studies~\cite{Thevenin02_dns_2d_3d,Thevenin05_kernel_dns_3d} will be investigated}.
\textcolor{black}{It will be shown in Sect.~\ref{sssec:flame_area_all} that these effects correlate with variations of the scalar dissipation term in the total flame kernel area balance equation.} By contrast, fluctuations of the area production term due to straining flow motion appear to be less important for \textcolor{black}{stochastic} growth of the young, engine-relevant flame kernels. \textcolor{black}{The predominant effect of curvature evolution on flame area dynamics will be confirmed} by complementary analysis of local regions in the planar-flame dataset in Sect.~\ref{sssec:flame_area_local}. \textcolor{black}{Different from early flame kernels, both the scalar dissipation term and the normal propagation term will be shown to exhibit pronounced variations in sections of the developed planar flame front. This difference can be explained by cancellation of run-to-run variations in normal propagation terms from both positively and negatively curved flame kernel regions, which does not occur in the developed planar flame.} \par
To identify important phenomena leading to enhanced or reduced global heat release rate, flame dynamics are studied by means of the scalar iso-surface area rate-of-change~\cite{Candel90_fsd_eq}. In this way, results from different flame geometries can be directly compared. The analysis is based on the reaction progress variable $\zeta$ defined by the solution of Eq.~(\ref{eq:c0_eq}). In a reference frame attached to the scalar iso-surface, the total time derivative of any quantity $\vartheta$ is given as:
\begin{equation}
\frac{\mathrm{D}_{\zeta} \left( \vartheta \right)}{\mathrm{D}_{\zeta}\left(t\right)} = 
\frac{\partial \vartheta}{\partial t} + u_i^{(\zeta)} \frac{\partial \vartheta}{\partial x_i} =
\frac{\partial \vartheta}{\partial t} + \left(u_i + \mathrm{s}_{\mathrm{d}}^{(\vartheta)} n_i \right) \frac{\partial \vartheta}{\partial x_i}.
\label{eq:sc_tot_d_dt}
\end{equation}
For $\vartheta=\zeta$, the total time derivative in the moving reference frame vanishes and Eq.~(\ref{eq:sc_tot_d_dt}) can be used to compute the iso-surface displacement speed $\mathrm{s}_{\mathrm{d}}$. \par
Since the flame structure has finite thickness with varying contributions of convection, reaction, and diffusion in addition to changing flow conditions across the flame, choosing one scalar iso-surface for flame-integral analysis may not be representative. Here, the time derivative of an averaged scalar iso-surface area~${\overline{A}_{\zeta,\Omega}=\int_{\Omega} \left|\nabla \zeta \right| \diff V}$ obtained by averaging over all reaction progress variable iso-surfaces~\cite{Boger98_gen_fsd}, is considered:
\begin{equation}
\frac{1}{\overline{A}_{\zeta,\Omega}} \frac{\mathrm{D}_{\zeta} \left( \overline{A}_{\zeta,\Omega} \right)}{\mathrm{D}_{\zeta}\left(t\right)} = 
\frac{\int_{\Omega}\left( 
a_{\mathrm{t}} +
\mathrm{s}_{\mathrm{rn}} \kappa -
D_{\mathrm{th}}\kappa^2
\right)\left|\nabla \zeta\right|\diff V}{\int_{\Omega}\left|\nabla \zeta\right|\diff V},
\label{eq:one_ov_A_dAdt}
\end{equation}
where the tangential strain rate is defined as:
\begin{equation}
a_{\mathrm{t}} = \frac{\partial u_k}{\partial x_k} - n_i \frac{\partial u_i}{\partial x_j} n_j = 
\frac{\partial u_{\mathrm{t}}}{\partial x_{\mathrm{t}}}.
\label{eq:aT_surf}
\end{equation}
Note that instead of using~$\mathrm{s}_{\mathrm{d}}$ in Eq.~(\ref{eq:one_ov_A_dAdt})~\cite{Candel90_fsd_eq}, the flame displacement contributions by normal propagation and tangential diffusion (cf.\ Eq.~(\ref{eq:sd_thin_rct_zones})) have been introduced. 
The second term in the r.h.s.\ of Eq.~(\ref{eq:one_ov_A_dAdt}) describes the total area change due to normal-propagation of curved flame regions (kinematic restoration, which could be source or sink for flame area depending on the sign of the curvature), while the third term is strictly negative and causes dissipation of flame area by diffusive scalar dissipation~\citep{Peters99_regime_diagr}.
%
\subsubsection{Integral Flame Surface Area Dynamics}
\label{sssec:flame_area_all}
\begin{figure}
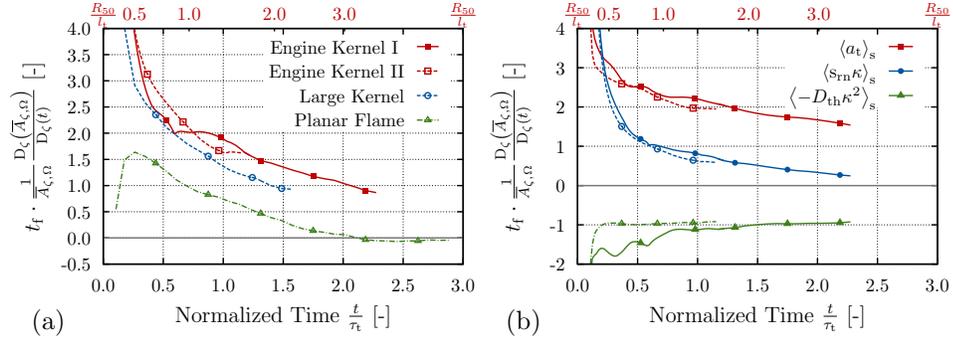
 
 \centering
 \begin{minipage}[b]{0.45\textwidth}
   \graphicspath{{./data/FLAME_KERNEL_DNS_02/181003_curv_strain_kin_restor_diss_divg_skew/}}
   \centering
   \input{template/change_font_10.tex} 
     \makebox[0pt][l]{\quad(a)}\scalebox{0.7}{\input{./data/FLAME_KERNEL_DNS_02/181003_curv_strain_kin_restor_diss_divg_skew/kernel_dns_mean_1ovA_dAdt_time_02_LOC01_03_P02_JFM_HALF.tex}}
   \input{template/change_font_12.tex} 
   \end{minipage}
  \begin{minipage}[b]{0.45\textwidth}
   \graphicspath{{./data/FLAME_KERNEL_DNS_02/181003_curv_strain_kin_restor_diss_divg_skew/}}
   \centering
  \input{template/change_font_10.tex} 
     \makebox[0pt][l]{\quad(b)}\scalebox{0.7}{\input{./data/FLAME_KERNEL_DNS_02/181003_curv_strain_kin_restor_diss_divg_skew/kernel_dns_mean_1ovA_dAdt_all_time_02_LOC01_JFM_HALF.tex}}
  \input{template/change_font_12.tex} 
   \end{minipage}
 \caption{Integral flame area rate-of-change for all flame configurations (a) \textcolor{black}{and individual terms (cf.\ Eq.~(\ref{eq:one_ov_A_dAdt})) for Engine Kernel~I (solid lines) and Engine Kernel~II (dashed lines) (b). The upper x-axes indicate the size of Engine Kernel~I.}}
 \label{fig:kernel_02_03_P02_dAdt_tot_curvSQ}
\end{figure}

In order to identify specific features of the small, engine-relevant flame kernel configuration, the integral flame surface area rates-of-change computed from all four DNS datasets are plotted in Fig.~\ref{fig:kernel_02_03_P02_dAdt_tot_curvSQ}\,(a). First, we focus on the dynamics of the planar flame, which was initialized from a laminar unstretched flame profile superimposed on the turbulent flow, without ignition heat source. Area production rapidly increases until a maximum is reached near ($t=0.25\cdot\tau_{\mathrm{t}}$), where dissipative effects begin to balance the production terms. At ($t=2.1\cdot\tau_{\mathrm{t}}$), the equilibrium between area production and dissipation is reached, i.e.\ the turbulent flame can be considered as developed. This result is in good agreement with Peters' estimate for flame brush development~\cite{Peters00_book}. \par
%
%
In contrast to the planar dataset, spherical flames with finite radius will always produce flame surface area. 
Another difference is caused by the ignition heat source applied to both, the large and the engine flame kernel configurations. Due to thermal expansion and accelerated flame propagation, area growth is initially very high. Until the end of the flame kernel simulations, net area production rates reduce, since the area dissipation mechanisms may develop similarly to the planar flame, and additionally, mean positive curvatures decrease due to flame growth, leading to less production of surface area (cf.\ Eq.~(\ref{eq:one_ov_A_dAdt})).\par
\textcolor{black}{Regarding the observed run-to-run variations in flame kernel surface area (cf.\ Fig.~\ref{fig:kernel_02_srcProg_Sigma_I0}\,(b)), note that Engine Kernel~I exhibits a plateau in net surface area production for ($0.6\tau_{\mathrm{t}}<t<0.9\tau_{\mathrm{t}}$), while the other flame configurations globally show strictly decreasing behavior. With the occurrence of the plateau, the total area rate-of-change of Engine Kernel~I becomes larger than in Engine Kernel~II, which reduces the difference in flame area between both kernel realizations (cf.\ Fig.~\ref{fig:kernel_02_srcProg_Sigma_I0}\,(b)). 
Similar deviations in global stretch rate were previously observed between flame kernel realizations in two-~\cite{Thevenin02_dns_2d_3d} and three-dimensional~\cite{Thevenin05_kernel_dns_3d} DNS. It will be shown in Sect.~\ref{sssec:flame_area_local} that also in local regions of the planar flame, net area rate-of-change fluctuates noticeably, even after the flame is on average fully developed.} 
Hereafter, the physical origin of this behavior will be further investigated since it can be relevant for~CCV in engines. \par
\textcolor{black}{For a deeper analysis, the production and dissipation terms of flame area are considered separately. To this end, all three terms in the r.h.s.\ of Eq.~(\ref{eq:one_ov_A_dAdt}) have been computed for both Engine Kernel~I and~II, and are plotted in Fig.~\ref{fig:kernel_02_03_P02_dAdt_tot_curvSQ}\,(b) as function of time. Overall, tangential strain is the major surface area production term, except for the very initial phase where large positive mean curvature leads to substantial area growth by flame normal propagation. The main dissipative effect under the present flame conditions can be attributed to scalar dissipation, which also applies to the developed planar flame as shown in Fig.~\mbox{S-10} of the supplementary material. It should be noted that the relative importance of kinematic restoration and scalar dissipation may depend on the ratio $l_{\mathrm{t}}/l_{\mathrm{f}}$, which is limited by the computationally affordable Reynolds number in the present work (cf.\ Tab.~\ref{tab:dns_params}). By comparison of both kernel realizations it seems that different flow conditions at the spark location are most prominently reflected in the dissipation term~${-D_{\mathrm{th}}\kappa^2}$, rather than the flame area production due to tangential strain~$a_{\mathrm{t}}$. This implies that the evolution of curvature plays a substantial role in flame area dynamics. However, one may expect variations in curvature to alter the normal propagation term~$\mathrm{s}_{\mathrm{rn}} \kappa$ as well, which is not evidenced by the results in Fig.~\ref{fig:kernel_02_03_P02_dAdt_tot_curvSQ}\,(b). This aspect will be revisited in the next section by comparison to results computed from local flame front segments of the planar flame.}\par
\subsubsection{Local Flame Surface Area Dynamics}
\label{sssec:flame_area_local}
%
\textcolor{black}{To investigate if the observed stochastic flame kernel area growth (cf.\ Sect.~\ref{sssec:flame_area_all}) can be fully attributed to the fact that the initially small flame area is statistically not representative, flame area dynamics in local regions of the planar flame with about the flame area of the small kernels will be considered hereafter.} 
For such analyses, the flow domain of the planar flame was divided into ($N_x\times N_y\times N_z=1\times 7\times 7$) subsets, where the $x$-direction is the mean propagation direction of the flame. This number of analysis volumes was chosen, since the instantaneous flame surface area of the planar flame at a characteristic time instant during flame kernel development ($t=0.6\cdot\tau_{\mathrm{t}}$) is approximately 49 times larger than the area of the small flame kernel. Expressed in integral length scales, each planar-flame subset contains a mean flame area of ($2\,l_{\mathrm{t}}\times 2\,l_{\mathrm{t}}$). 
%
%
\textcolor{black}{In the following, it will be shown that the crucial role of flame curvature for flame area dynamics during early flame kernel development translates to local regions of the developed planar flame, but appears through different mechanisms.} \par
\begin{figure}
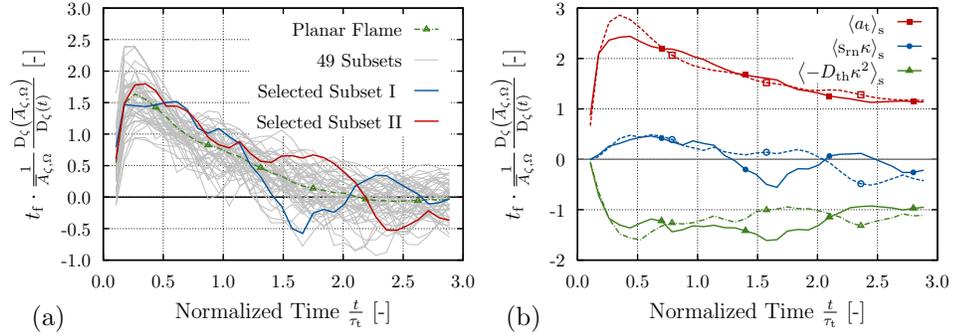
 
 \centering
 \begin{minipage}[b]{0.45\textwidth}
   \graphicspath{{./data/FLAME_PLANAR_DNS_02/180923_curv_strain_kin_restor_skew_curvEQ_subdomains7x7/}}
   \centering
 \input{template/change_font_10.tex} 
  \makebox[0pt][l]{\quad(a)}\scalebox{0.7}{\input{./data/FLAME_PLANAR_DNS_02/180923_curv_strain_kin_restor_skew_curvEQ_subdomains7x7/flame_kernel_Ka_stretch_49_subsets_fast_P02_JFM_HALF.tex}}
 \input{template/change_font_12.tex} 
   \end{minipage}
  \begin{minipage}[b]{0.45\textwidth}
   \graphicspath{{./data/FLAME_PLANAR_DNS_02/181003_curv_strain_kin_restor_diss_divg_skew_190712_subdomains7x7/}}
   \centering
 \input{template/change_font_10.tex} 
  \makebox[0pt][l]{\quad(b)}\scalebox{0.7}{\input{./data/FLAME_PLANAR_DNS_02/181003_curv_strain_kin_restor_diss_divg_skew_190712_subdomains7x7/kernel_dns_mean_1ovA_dAdt_all_time_P02sub_07_34_JFM_HALF.tex}}
 \input{template/change_font_12.tex} 
   \end{minipage}
 \caption{Flame area rate-of-change in local regions of the planar flame (a) \textcolor{black}{and individual terms (cf.\ Eq.~(\ref{eq:one_ov_A_dAdt})) for Subset~I (solid lines) and Subset~II (dashed lines) (b)}.}
 \label{fig:kernel_02_03_P02_dAdt_tot_curvSQ_sub}
\end{figure}

In Fig.~\ref{fig:kernel_02_03_P02_dAdt_tot_curvSQ_sub}\,(a), the time evolution of flame area rate-of-change in each local analysis segment of the planar flame is compared to the total flame area change, analogously to Fig.~\ref{fig:kernel_02_03_P02_dAdt_tot_curvSQ}\,(a). It can be concluded that locally, the flame area production/dissipation balance fluctuates noticeably, even after the turbulent flame is on average fully developed. Similar to the analysis in Sect.~\ref{sssec:flame_area_all}, the origin of the fluctuations will be briefly discussed, focussing on the minimum/maximum in area rate-of-change of the two selected subsets during ($1.25\tau_{\mathrm{t}}<t<2.1\tau_{\mathrm{t}}$). 
%
%
\textcolor{black}{According to Fig.~\ref{fig:kernel_02_03_P02_dAdt_tot_curvSQ_sub}\,(b), local variations in flame area production by tangential strain are small during this time interval. However, pronounced variations in both curvature-dependent terms occur, which cause the temporal variations in net flame area dynamics of the selected flame front segments. This is different from the engine-relevant flame kernel configuration, where no significant run-to-run variations in the normal propagation term were observed. Note that there seems to be some correlation between the strictly negative scalar dissipation effect~${-D_{\mathrm{th}}\kappa^2}$, and the normal propagation term~$\mathrm{s}_{\mathrm{rn}} \kappa$, which mainly follows the sign of curvature. Hence, the temporal variations in local net area dynamics of the developed planar flame might be attributed to variations in negatively curved regions.} \par
\begin{figure}
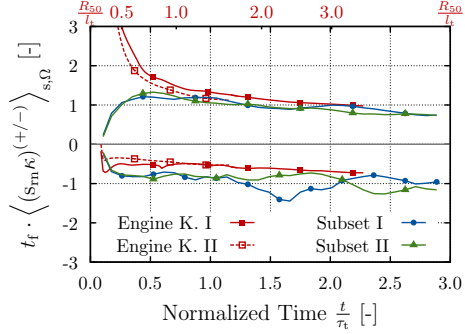
 
 \centering
 \begin{minipage}[b]{0.45\textwidth}
   \graphicspath{{./data/FLAME_KERNEL_DNS_02/181003_curv_strain_kin_restor_diss_divg_skew/}}
   \centering
\input{template/change_font_10.tex} 
\scalebox{0.7}{\input{./data/FLAME_KERNEL_DNS_02/181003_curv_strain_kin_restor_diss_divg_skew/kernel_dns_mean_1ovA_dAdt_PROP_plus_minus_time_02_LOC01_P02sub_JFM_HALF.tex}}
\input{template/change_font_12.tex} 
   \end{minipage}
 \caption{\textcolor{black}{Flame area rate-of-change due to normal propagation split into contributions from positively and negatively curved flame regions. The upper x-axis indicates the size of Engine Kernel~I.}}
 \label{fig:kernel_02_P02_dAdt_terms_sub_prop_plus_minus}
\end{figure}

\textcolor{black}{To clarify this difference between small flame kernels and the developed flame, the normal propagation term has been split into contributions from positively and negatively curved flame regions. The results depicted in Fig.~\ref{fig:kernel_02_P02_dAdt_terms_sub_prop_plus_minus} confirm that in the planar flame, local variations in flame area rate-of-change can be attributed to the dissipative (negative) kinematic restoration effect. By contrast, differences between flame kernel realizations occur irrespectively of the curvature sign, which mostly cancel out when the net effect of flame normal propagation on flame area dynamics, i.e.\ $\mathrm{s}_{\mathrm{rn}} \kappa$, is considered. Since the evolution of~$\overline{\mathrm{I}}_{0,\mathrm{rn}}$ was shown be very similar for both Engine Kernel~I and~II (cf.\ Fig.~\ref{fig:kernel_02_mean_I0}), different flow conditions at the spark location of Engine Kernel~I seem to result in larger curvature magnitudes, i.e.\ different flame geometry in both positively and negatively curved flame regions. These run-to-run variations in curvature evolution due to flame kernel/turbulence interactions will be investigated in a future study.} \par

\section*{Conclusions}
\label{sec:conclusions}
\textcolor{black}{Since the reduction of CCV is crucial for improving SI engine efficiency and emissions, an understanding of the root causes of the known CCV correlation with early flame kernel development is desirable. In this work, a DNS database has been carefully designed to represent engine part load conditions while allowing for systematic parametric studies.
The analysis has been focussed on quantities that are of particular interest for modeling the integral heat release rate, i.e.\ mean flame displacement speed and flame area (or flame brush) evolution.
Under the present fully homogeneous, engine-relevant conditions at unity Lewis number, the following conclusions can be drawn with respect to the flame structure and flame displacement speed:}
\begin{itemize}
\item Turbulent perturbations of the preheat zone develop within short time ($t<0.25\cdot\tau_{\mathrm{t}}$).
\item On average, no flame thickening due to turbulent mixing is found, \textcolor{black}{despite a Karlovitz number of $\mathrm{Ka}\approx10$}.
\item After ignition effects have decayed, the mean normal-displacement speed is identical 
to the laminar burning velocity of an unstretched premixed flame ($\overline{\mathrm{I}}_{0,\mathrm{rn}}=1$).
\end{itemize}
These observations are 
equally valid for more realistic Reynolds numbers, where we expect similar Karlovitz but higher Damk{\"o}hler numbers. \par
\textcolor{black}{Regarding run-to-run variations in the global heat release rate during early flame kernel development, variations in flame geometry have been identified as the primary cause under the present flame conditions. }
The analysis of flame area rate-of-change has shown the following:
\begin{itemize}
\item The engine-relevant flame kernel configuration exhibits run-to-run variations in net flame area production, comparable to previous studies~\cite{Thevenin02_dns_2d_3d,Thevenin05_kernel_dns_3d}.
\item \textcolor{black}{Complementary analyses of local flame front segments of the planar flame with about the flame area of the small kernels have shown that temporal variations in integrated flame area rate-of-change occur in developed flames as well.}
\item \textcolor{black}{In both flame configurations, the variations are mainly caused by the curvature-dependent terms in the flame area balance equation, not by the production terms due to strain.}
\item \textcolor{black}{Local variations in flame area of the developed flame can be attributed to negatively curved flame front portions. By contrast, run-to-run variations in total flame kernel surface area are equally caused by differences in curvature evolution in both positively and negatively curved regions due to different flow conditions at the two investigated spark locations.}
\end{itemize}
While the present findings on flame kernel area evolution highlight the importance of curvature dynamics during flame kernel development, differential diffusion and spark ignition effects additionally play an important role\textcolor{black}{~\cite{Falkenstein19_kernel_Le_I_cnf,Falkenstein19_kernel_Le_II_cnf}, which require further consideration particularly under flame conditions representative for off-stoichiometric and EGR-diluted engine operation. Finally, it should be noted that the main effects discussed in the present study were generally also observed in equivalent non-unity-Lewis-number flames, but the quantitative extent depends on the actual flow conditions and respective flame response.}
%
\section*{Acknowledgement}
The authors from RWTH Aachen University gratefully acknowledge partial funding by Honda R\&D and by the Deutsche Forschungsgemeinschaft (DFG, German Research Foundation) under Germany's Excellence Strategy - Exzellenzcluster 2186 `The Fuel Science Center' ID: 390919832.\par
The authors gratefully acknowledge the Gauss Centre for Supercomputing~e.V. (www.gauss-centre.eu) for funding this project by providing computing time on the GCS Supercomputer Super-MUC at Leibniz Supercomputing Centre (LRZ, www.lrz.de).\par
Data analyses were performed with computing resources granted by RWTH Aachen University under project thes0373. \par
S.K. gratefully acknowledges financial support from the National Research Foundation of Korea (NRF) grant by the Korea government (MSIP) (No. 2017R1A2B3008273). \par

\appendix

\bibliography{./references/flame_kernel_01,./references/les4ice16.bib,./references/rezchikova.bib, ./references/03_b_Publikationen_anderer.bib}

\end{document}